\documentclass[12pt]{article}
\def\be{\begin{equation}}
\def\eq{\end{equation}}
\usepackage{amsmath}
\usepackage{amssymb}

\def\cc {\c c}
\def\[{\begin{equation}}
\def\]{\end{equation}}

\begin{document}
\title{An English translation of Bertrand's theorem }
\author{F C Santos
\footnote{e-mail: filadelf@if.ufrj.br}\\ V Soares\footnote{e-mail: vsoares@if.ufrj.br} \\ A C Tort
\footnote{e-mail: tort@if.ufrj.br.}\\
Instituto de F\'{\i}sica
\\
Universidade Federal do Rio de Janeiro\\
Caixa Postal 68.528; CEP 21941-972 Rio de Janeiro, Brazil}
\maketitle

\begin{abstract}
A beautiful theorem due to J. L. F. Bertrand concerning the laws of attraction that admit bounded closed orbits for arbitrarily chosen initial conditions is translated from French into English.
\end{abstract}
\vskip 0.25cm
PACS: 45.50.Dd; 45.00.Pk
\section{Introduction}
In 1873, Joseph Louis Fran\cc ois Bertrand  (1822-1900) \cite{bertrand} published a short but important paper in which he proved that there are two central fields only for which all bounded orbits are closed, namely, the isotropic harmonic oscillator law and Newton's universal gravitation law, that Bertrand calls \lq la loi de la nature\rq   \,(the law of nature). Because of this additional symmetry it is no wonder that the most essential properties of these two fields were studied by Newton himself who discusses them in the Proposition X and in the Proposition  XI of Book I of his \textit{Principia} \cite{Newton}. Newton shows that both fields give rise to an elliptical orbit with the difference that in the first case the force is directed towards the geometrical centre of the ellipse and in the second case the force is directed to one of the foci. Bertrand's paper appeared in the \textit{Comptes Rendus} of the Acad\'emie des Sciences de Paris where the \textit{m\'emoires} and communications of the members and correspondent members of that French science academy were published.  The academic session at which Bertrand presented his paper took place on Monday 20th October 1873 under the presidency of Mr. de Quatrefages. Bertrand's result, also known as Bertrand's theorem, continues to fascinate old and new generations of physicists interested in classical mechanics and unsurprisingly papers devoted to it continue to be produced and published. Bertrand's proof is concise and elegant and contrary to what one may be led to think by a number of perturbative demonstrations that can be found in modern literature, textbooks and papers on the subject, is fully non-perturbative. Giving the continual interest on this theorem and the fact that English is the \textit{lingua franca} of modern science, the present authors think that an English version of Bertrand's original proof may be of some value. 

\section{The translation: {\small ANALYTICAL MECHANICS}. -- \textit{A theorem relative to the motion of a point pulled towards a fixed centre; by Mr.} \textbf{J. Bertrand. } }
The planetary orbits are closed curves; this is the main cause of the stability of our system, and this important circumstance stems from the law of attraction which, whatever the initial circumstances, makes each celestial body which is not expelled from our system follow the circumference of an ellipse. Until now it was not observed that the Newtonian law of attraction is the only one that fulfills this condition.

Among the laws of attraction that assume to be null the action at an infinite distance, that of nature is the only one for which a mobile body \textit{arbitrarily} launched with a speed less that a certain limit and pulled towards a fixed centre, describes necessarily a closed curve about this centre.  All laws of attraction \textit{allow} closed orbits, but the law of nature is the only one that \textit{imposes} them.

We prove this theorem in the following way: let $\varphi (r) $ be the attraction exerted at a distance $r$ on the molecule\footnote{In the original French manuscript, \textit{mol\'ecule}. Bertrand is certainly referring to a particle. } and directed to the centre of the attraction that we will take as the origin of the coordinates.  Denoting by $r$ e $\theta$ the two polar coordinates of the mobile body, we have by virtue of a well known formula,

$$ \varphi (r) = \frac{k^2}{r^2}\left(  \frac{1}{r} + \frac{d^2 \frac{1}{r}}{d\theta^2}\right) , $$
and, setting $\frac{1}{r}=z $,

\be
r^2\varphi (r) =\psi (z),
\eq

$$ \frac{d^2 z}{d\theta^2}+z-\frac{1}{k^2}\psi (z) =0 . $$
Let us multiply both members by $2\,dz$ and let us integrate setting 
\be
2 \int \psi (z)\,dz =\varpi (z),
\eq
we will have 

$$ \left( \frac{ dz}{d\theta} \right)^2+z^2-\frac{1}{k^2}\varpi (z) - h=0 ,$$ $h$ being a constant.

From this one deduces that

$$ d\theta = \pm \frac{dz}{\sqrt{h + \frac{1}{k^2}\varpi (z) - z^2} }. $$

If the curve represented by the equation that ties $z$ to $\theta$ is closed, the value of $z$ will have maxima and minima for which $dz/d\theta $ will be null and the corresponding vector radii, normal to the trajectory, will  necessarily be axes of symmetry for it. Now when a curve admits two axes of symmetry, the necessary and sufficient condition for it to be closed is that its angle be commensurable with $\pi$.  Therefore, if $\alpha$ and $\beta$ represent  a minimum of $z$ and the maximum that follows it, the condition required is expressed by the equation

\be
m\pi=\int_\alpha^\beta \, \frac{dz}{\sqrt{h + \frac{1}{k^2}\varpi (z) - z^2} } ,
\eq
where $m$ denotes a commensurable number. This equation must hold whatever $h$ and $k$ might be, and consequently, the limits $\alpha$ and $\beta$ that depend on them. 

One has

$$h + \frac{1}{k^2} \varpi (\alpha ) -\alpha^2=0 ,$$

$$h + \frac{1}{k^2} \varpi (\beta ) -\beta^2=0 ;$$
consequently
$$ \frac{1}{k^2}=\frac{\beta^2-\alpha^2}{\varpi (\beta)-\varpi (\alpha) } , $$

$$ h =\frac{\alpha^2\varpi (\beta) -\beta^2 \varpi (\alpha )}{\varpi (\beta)-\varpi (\alpha) } , $$
and equation (3) becomes

\be
m\pi= \int_\alpha^\beta \, \frac{dz\,\sqrt{\varpi (\beta)-\varpi (\alpha) } }{\sqrt{\alpha^2\varpi (\beta) -\beta^2 \varpi (\alpha) +\left( \beta^2-\alpha^2 \right) \varpi (z) -z^3[\varpi (\beta) -\varpi (\alpha )] }} .
\eq
The function $\varpi (z) $ must be such that this equation holds for all values of $\alpha$ e $\beta$. Moreover, the commensurable number $m$ must be constant, for if it were to vary from one orbit to another one, an infinitely small variation of the initial conditions would bring forth a finite variation of the number and the disposition of the axes of symmetry of the trajectory.

Assume that $\alpha$ and $\beta$ differ infinitesimally; let

$$ \beta=\alpha + u , $$
$z$ staying included between $ \alpha $ and $\beta$, we can set

$$ z=\alpha +\gamma ,$$
and $\gamma$ will be, just as $u$, infinitely small. Neglecting the  infinitely small of second order we will have

$$ \sqrt{\varpi (\beta ) -\varpi (\alpha ) } = \sqrt{u\,\varpi^{\,\prime} (\alpha)} . $$
In the expression under the radical sign in the denominator of the integral (4)
the infinitely small of first order reduce to zero, and the same happens with those of second; it is those of third that are necessary to keep, and neglecting the infinitely small of fourth order one has
\begin{eqnarray*}
& & \alpha^2  \varpi (\beta) - \beta^2 \varpi (\alpha) +\left( \beta^2-\alpha^2 \right) \varpi (z) -z^3[\varpi (\beta) -\varpi (\alpha )] \nonumber \\ 
& = & [ \varpi^{\,\prime} (\alpha) -\alpha \varpi^{\,\prime\prime} (\alpha) ]  \left(u^2\gamma--u \gamma^2 \right) .
\end{eqnarray*}
Equation (4) becomes

$$ m\pi = \int_0^u \, \frac{d\gamma \,\sqrt{\varpi^{\,\prime} (\alpha) }}{\sqrt{ \varpi^{\,\prime} (\alpha) -\alpha \varpi^{\,\prime\prime} (\alpha) } \,\sqrt{u\gamma-\gamma^2} } , $$
that is, performing the integration and suppressing common factors

$$ m= \sqrt{\frac{\varpi^{\,\prime} (\alpha ) }{\varpi^{\,\prime} (\alpha) -\alpha \varpi^{\,\prime\prime} (\alpha)} } ,$$
or

$$\left(1-m^2\right)\varpi^{\,\prime} (\alpha) +m^2\alpha \varpi^{\,\prime\prime} (\alpha) =0 .$$
From this one deduces that

$$ \varpi^{\,\prime} (\alpha)= \frac{A}{\alpha^{1/m^2 -1 }} ,$$

$$\varpi (\alpha) = A\,\frac{\alpha^{2-1/m^2}}{2-\frac{1}{m^2}} + B, $$
$A$ and $B$ denoting constants. 

From the assumed relations between the functions $\varpi$, $\psi$ e $\varphi$ it follows that

$$ \psi (z) = \frac{A}{2 \, z^{1/m^2-1} } , $$

$$ \varphi (r) = \frac{A}{2}\,r^{1/m^2-3} . $$
Such is the only possible law of attraction, $m$ here denoting any commensurable number; but from this it does not follow that it fulfills all the conditions of the proposition for any $m$. In fact, one must have for values of $\alpha$ and $\beta$,
\begin{equation*}
m\pi=\int_\alpha^\beta\, \frac{dz\,\sqrt{\frac{1}{\beta^{1/m^2-2} } -\frac{1}{\alpha^{1/m^2-2} }} }{\frac{\alpha^2}{\beta^{1/m^2-2}} - \frac{\beta^2}{\alpha^{1/m^2-2} } + (\beta^2-\alpha^2)\frac{1}{z^{1/m^2-2}} -z^2\left( \frac{1}{\beta^{1/m^2-2}}-\frac{1}{\alpha^{1/m^2-2}}\right)} . \hskip 0.50cm \mbox{(6)}
\end{equation*}

\footnote{We have kept the original enumeration of the equations and for this reason there is no equation (5).} Let us assume initially $1/m^2-2 $ negative; let us set $\alpha=0$, $\beta=1$, the equation becomes

$$ m\pi = \int_0^1 \frac{dz}{\sqrt{ \frac{1}{z^{1/m^2-2}} -z^2} }=\int_0^1\,\frac{z^{1/(2m^2)-1}\,dz}{1-z^{1/m^2}},$$
and equation (6) yields

$$ m\pi = m^2 \pi ,$$

$$m=1 .$$
The corresponding law of attraction is

$$ \varphi (r) = \frac{A}{r^2} .$$

If one assumes $1/m^2-2$ positive, equation (6) for $\alpha=1$, $\beta=0$,

$$ m\pi=\int_0^1\, \frac{dz}{\sqrt{1-z^2}}=\frac{\pi}{2} .$$
From this it follows that $m=1/2$, and the corresponding law of attraction is 

$$ \varphi (r) = A\,r . $$

Only two laws therefore fulfill the required conditions, that of nature, by which the closed orbit has only one symmetry axis passing through the centre of action, and the attraction proportional to the distance, by which there are two. 

Our illustrious correspondent Mr. Chebychev\footnote{In the original manuscript, Tchebychef.}, to whom I have communicated the precedent demonstration, judiciously made me remark that the theorem, useless nowadays for the so perfect theory of the planets, may be invoked in a useful way in order to extend to the double stars the newtonian laws of attraction. 


\section{Final remarks}
Until the submission for publication of this translation, Bertrand's original paper could be found at the following address http://gallica.bnf.fr/ark:/12148/bpt6k3034n.



\end{document}